\documentclass[aps,prb,reprint]{revtex4-1}
\usepackage{graphicx}
\usepackage{mathrsfs}
\usepackage{bm}
\usepackage{amsmath}
\usepackage{dcolumn}
\usepackage{epstopdf}
\usepackage{dsfont}
\usepackage{amssymb}
\usepackage{tabularx}
\usepackage{array}
\usepackage{color}
\usepackage{braket}
\usepackage[colorlinks=true, letterpaper=true, pdfstartview=FitV, linkcolor=blue, citecolor=blue, urlcolor=blue]{hyperref}

\begin{document}

\title{Enhanced magneto-optical response due to the flat band in nanoribbons
made from the $\alpha-T_3$ lattice}
\author{Yan-Ru Chen$^{1,2}$}
\author{Yong Xu$^{2}$}
\author{Jun Wang$^3$}
\author{Jun-Feng Liu$^{1,2}$}
\email{phjfliu@gzhu.edu.cn}
\author{Zhongshui Ma$^{4,5}$}
\email{mazs@pku.edu.cn}

\affiliation{$^1$Department of Physics, School of Physics and Electronic Engineering, Guangzhou University, Guangzhou
510006, China}
\affiliation{$^2$Department of Physics, Southern University of Science and Technology,
Shenzhen 518055, China}
\affiliation{$^3$Department of Physics, Southeast
University, Nanjing 210096, China}
\affiliation{$^4$School of Physics, Peking
University, Beijing 100871, China}
\affiliation{$^5$Collaborative Innovation Center of Quantum Matter, Beijing,
100871, China}

\begin{abstract}
We study the optical response of nanoribbons made from the $\alpha-T_3$
lattice under a weak magnetic field in the terahertz to far-infrared regime.
It is found that the magnetic field can open a gap in the band structure and
induce a new absorption peak with much reduced frequency in metallic
armchair ribbons and a class of zigzag ribbons with particular boundaries.
This tunable magneto-optical modulation effect is attributed to the
interband transitions between the flat band and the propagating bands. By
contrast, this magnetic modulation of gap opening and optical conductance is
much weaker in metallic armchair graphene ribbons (the case of $\alpha=0$)
in which the flat band is absent. The enhancement in the $\alpha-T_3$ model
is analytically investigated and explained within the perturbation theory
for metallic armchair ribbons. The magnetic field induced valley degeneracy
lifting and valley splitting of the absorption peak are also discussed in
the case of zigzag ribbons. These findings pave the way for magneto-optics
devices based on the $\alpha-T_3$ model materials.
\end{abstract}

\maketitle

\section{Introduction}

\begin{figure}[tbp]
\begin{center}
\includegraphics[bb=11 8 336 366, width=3.415in]{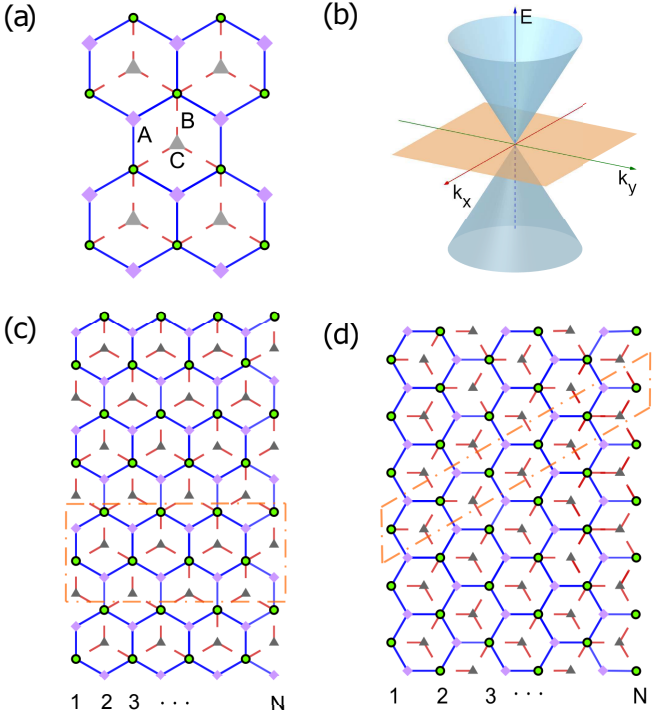}
\end{center}
\caption{(a) The lattice of the $\protect\alpha-T_3$ model. (b) The
low-energy dispersion at a single $K$ point. (c) An armchair ribbon from the
$\protect\alpha-T_3$ lattice with the width $N$. (d) A B-B edged zigzag
ribbon from the $\protect\alpha-T_3$ lattice with the width $N$ where B
atoms terminates at both the left and right edges.}
\label{model}
\end{figure}

Graphene, as a two-dimensional sheet of carbon atoms arranged on a honeycomb
lattice, has achieved a great success in both fundamental physics and
applications due to its exotic electronic properties \cite{gra1,gra2,gra3}.
Recently, a modified lattice -- the $\alpha-T_3$ lattice, is attracting more
and more attention as an interpolation between the honeycomb lattice of
graphene and the dice lattice. As shown in Fig. \ref{model}(a), the dice or $%
T_3$ lattice is obtained by coupling one of the two inequivalent sites of
the honeycomb lattice to an additional atom located at the center of each
hexagon \cite{vidal98,vidal01,dora11}. The dice lattice could be
experimentally realized by growing a trilayer structure of cubic lattices
such as SrTiO$_3$/SrIrO$_3$/SrTiO$_3$ in the (111) direction \cite{wang11}
or by confining cold atoms to an optical lattice \cite{bercioux09}. The
low-energy quasiparticle in the dice lattice is described by the
pseudospin-1 Dirac-Weyl equation \cite{dora11,bercioux09}. The spectrum
contains a pair of linear Dirac cones and an additional dispersionless flat
band that cuts through the Dirac points (see Fig. \ref{model}(b)). The $%
\alpha-T_3$ lattice interpolates between graphene ($\alpha=0$) and the dice
lattice ($\alpha=1$) via a parameter $\alpha $ that describes the strength
of the coupling between the honeycomb lattice and the central hub site.
Recently, a 2D model for Hg$_{1-x}$Cd$_x$Te at critical doping has been
shown to map onto the $\alpha-T_3$ model with an intermediate parameter $%
\alpha=1/\sqrt{3}$ \cite{malcolm15}. The $\alpha-T_3$ model has also been
generalized to include additional terms and variations in its Hamiltonian
\cite{piechon15}. And the properties of general pseudospin $S$ lattices have
also been extensively studied \cite{dora11,malcolm14,lan11}.

The $\alpha$-dependent Berry phase \cite{raoux14,louvet15,xiao10} in the $%
\alpha-T_3 $ model results in unusual electronic properties such as
unconventional quantum Hall effect\cite{yong17,biswas16}, supper-Klein
tunneling \cite{daniel11,xing10,illes17,coss17}, minimal conductivity \cite%
{louvet15}, orbital magnetic response \cite{raoux14}, Wiess oscillations
\cite{sk17} etc. Additionally, the flat band also plays an important role in
the transport. Although the flat band itself has zero conductivity due to
the zero group velocity, the interplay between the flat band and the
propagating bands is predicted to induce a diverging dc conductivity in the
presence of disorders \cite{vigh13}, or enhance the resulting current in a
nonequilibrium situation \cite{wangcz17}. Additionally, the flat band have
also attracted much attention for its nontrivial topology \cite%
{fengliu13,yamada16,fengliu18,peotta15,wenxg11,sarma11,mudry11} and
interaction effect \cite{peotta15,wenxg11,sarma11,mudry11,fengliu11}. It is
noticeable that the first-principles calculations implied that the flat band
can exist in a realistic material \cite{fengliu13,yamada16,fengliu18}.

The optical and magneto-optical spectroscopy can be used to probe the
underlying electronic structure and as well as design optoelectronic
devices. For graphene, the optical and magneto-optical conductivities have
been extensively studied for both infinite sheet \cite{mazs08} and
nanoribbon geometry \cite%
{prezzi08,hsu07,liu08,zhang09,zhou081,zhou082,have18}. The optical response
at selective frequencies can be enhanced with the use of nanoribbons in
graphene. For the $\alpha -T_{3}$ lattice, the optical \cite{illes15} and
magneto-optical conductivities \cite{malcolm14,illes16,aron17} have been
studied for infinite sheet. But the optical conductance of nanoribbons from
the $\alpha -T_{3}$ lattice (see Figs. \ref{model}(c) and \ref{model}(d))
has not been investigated. The effect of a perpendicular magnetic field on
the electronic structure and optical conductance of $\alpha -T_{3}$
nanoribbons is also untouched. It has been shown that the magnetic filed can
open a gap and induce an absorption peak in metallic armchair graphene
nanoribbons \cite{liu08}. It is natural to raise the question of how this
magnetic modulation effect on electronic and optical properties of
nanoribbons evolves with varying parameter $\alpha $ in the $\alpha -T_{3}$
lattice.

In this work, we investigate systematically the magnetic modulation effect
on electronic and optical properties of nanoribbons made from the $%
\alpha-T_3 $ lattice. We show that the magnetic field can open a gap and
induce an absorption peak in both the metallic armchair ribbons and a
special type of zigzag ribbons. This magnetic modulation effect is
remarkably enhanced in contrast to that in graphene due to the interband
transitions between the flat band and the propagating bands, which makes the
optical response much more sensitive to the applied magnetic field. We also
present analytical discussions to explain this enhanced magnetic modulation
effect in the $\alpha-T_3$ lattice. Without the magnetic field, the low
frequency conductivity is usually zero and the first response peak appears
at a frequency corresponding to the allowed transition with the lowest
energy gap. We show that the large threshold frequency that prevented
terahertz and far-infrared (FIR) response in ribbons can be reduced to a
much lower value by a magnetic field, making these ribbons active in the
terahertz and FIR regime. The applied magnetic field is much weaker than
that needed in graphene. Therefore controlled terahertz radiation can be
achieved much more easily in nanoribbons made from the $\alpha-T_3$ lattice
through the applied magnetic field.

The rest of the paper is organized as follows. In Sec. II, we present the
model and the Kubo formula for the calculation of the optical conductivity.
The numerical results and discussions on electronic structure and optical
conductivity will be presented for armchair ribbons in Sec. III, and for
zigzag ribbons in Sec. IV. Finally, conclusion remarks are given in Sec. V.

\section{Model and Kubo formula}

In the $\alpha -T_{3}$ lattice shown in Fig. \ref{model}(a), sites A and B
form honeycomb lattice, and site C at the center of the hexagons. The
hopping energy between sites A and B is $t_{1}=t\cos {\varphi }$, and the
hopping between sites B and C is $t_{2}=t\sin {\varphi }_{.}$ The parameter $%
\alpha $ is defined as $\alpha =t_{2}/t_{1}=\tan {\varphi }$. And the
hopping between sites A and C is not permitted. This $\alpha -T_{3}$ model
interpolates between the honeycomb lattice of graphene and the dice lattice
via the parameter $\alpha$. Under a perpendicular magnetic field, the
tight-binding Hamiltonian on the basis $(\psi _{A},\psi _{B},\psi _{C})^{T}$
is given by
\begin{equation}
H=\sum_{<i,j>,<j,k>}(t\cos {\varphi }e^{i\gamma _{ij}}a_{i}^{\dag
}b_{j}+t\sin {\varphi }e^{i\gamma _{jk}}b_{j}^{\dag }c_{k}+\text{H.c.})
\label{hamil1}
\end{equation}%
where $\varphi =\arctan {\alpha }$, $a^{\dag }$, $b^{\dag }$, and $c^{\dag }$
($a$, $b$, $c$) are creation (annihilation) operators at sites A, B, C
respectively. Here $\gamma _{ij(jk)}=(2\pi /\phi _{0})\int_{i(j)}^{j(k)}%
\mathbf{A}\cdot d\mathbf{l}$ is the magnetic Peierls phase, with $\phi
_{0}=hc/e$ being the magnetic flux quantum. The magnetic field in the $z$
direction is described by the vector potential $\mathbf{A=}Bx\hat{y}$ in the
Landau gauge. In the nanoribbon geometry shown in Figs. \ref{model}(c) and %
\ref{model}(d), the Hamiltonian can be constructed on the basis of the
supercell from the Harper equation.

Without the magnetic field, the continuum Hamiltonian of the $\alpha -T_{3}$
model can be written as
\begin{equation}
H_{0}=%
\begin{pmatrix}
0 & \cos {\varphi }f(\boldsymbol{k}) & 0 \\
\cos {\varphi }f^{\ast }(\boldsymbol{k}) & 0 & \sin {\varphi }f(\boldsymbol{k%
}) \\
0 & \sin {\varphi }f^{\ast }(\boldsymbol{k}) & 0%
\end{pmatrix}%
,  \label{hamil2}
\end{equation}%
where $f(\boldsymbol{k})=t(1+e^{-i\boldsymbol{k}\cdot \boldsymbol{a}%
_{1}}+e^{-i\boldsymbol{k}\cdot \boldsymbol{a}_{2}})=|f(\boldsymbol{k}%
)|e^{i\theta _{k}}$ with $\boldsymbol{k}$ the momentum and $\theta _{k}$
being the complex angle of $f(\boldsymbol{k})$, $\boldsymbol{a}_{1}=(-\frac{%
\sqrt{3}}{2},\frac{3}{2})a$, $\boldsymbol{a}_{2}=(\frac{\sqrt{3}}{2},\frac{3%
}{2})a$. The eigenvalues can be obtained as $E_{0}=0$ and $E_{\pm }=\pm |f(%
\boldsymbol{k})|$. The corresponding eigenfunctions are
\begin{equation}
\xi _{0}=%
\begin{pmatrix}
\sin {\varphi }e^{i\theta _{k}} \\
0 \\
-\cos {\varphi }e^{-i\theta _{k}}%
\end{pmatrix}%
,\quad \xi _{\pm }=\frac{1}{\sqrt{2}}%
\begin{pmatrix}
\cos {\varphi }e^{i\theta _{k}} \\
\pm 1 \\
\sin {\varphi }e^{-i\theta _{k}}%
\end{pmatrix}%
\end{equation}%
with the $\xi _{0}$ denoting the flat band, $\xi _{\pm }$ denoting the
conduction ($+$) and valence ($-$) bands. The full wave function reads $\psi
=\xi e^{i(k_{x}x+k_{y}y)}$.

To consider nanoribbons, we construct the Harper eqation from the lattice
Hamiltonian. For armchair ribbons, the Harper equation reads
\begin{equation}
E\psi _{m}=B_{m-1}\psi _{m-1}+D_{m}\psi _{m}+B_{m}\psi _{m+1}
\end{equation}%
where $m$ is the cell index in a supercell as shown in Fig. \ref{model}(c), $%
\psi _{m}=%
\begin{pmatrix}
\psi _{Am} & \psi _{Bm} & \psi _{Cm}%
\end{pmatrix}%
^{T}$. On the basis $\psi =%
\begin{pmatrix}
\psi _{1} & \psi _{2} & \psi _{3} & \cdots & \psi _{N}%
\end{pmatrix}%
^{T}$, we can construct the Hamiltonian of an armchair nanoribbon with width
$N$ as follows
\begin{equation}
H_{arm}=%
\begin{pmatrix}
D_{1} & B_{1} & 0 & \cdots & 0 \\
B_{1} & D_{2} & B_{2} & \cdots & 0 \\
0 & B_{2} & D_{3} & \cdots & 0 \\
\cdots & \cdots & \cdots & \cdots & \cdots \\
0 & 0 & 0 & \cdots & D_{N}%
\end{pmatrix}%
,
\end{equation}%
where
\begin{align}
& B_{m}=%
\begin{pmatrix}
0 & \cos {\varphi }g_{1m} & 0 \\
\cos {\varphi }g_{1m}^{\ast } & 0 & \sin {\varphi }g_{1m} \\
0 & \sin {\varphi }g_{1m}^{\ast } & 0%
\end{pmatrix}%
, \\
& D_{m}=%
\begin{pmatrix}
0 & \cos {\varphi }g_{2m} & 0 \\
\cos {\varphi }g_{2m}^{\ast } & 0 & \sin {\varphi }g_{2m} \\
0 & \sin {\varphi }g_{2m}^{\ast } & 0%
\end{pmatrix}%
.
\end{align}%
Here $g_{1m}=t\exp {[i\frac{\pi }{3}f(m^{\prime }+\frac{1}{2})+\frac{a}{2}%
k_{y}]}$ and $g_{2m}=t\exp {[-i(\frac{2}{3}\pi fm^{\prime }+k_{y}a)]}$, with
$m^{\prime }=m-(N+1)/2$ and $f=3\sqrt{3}Ba^{2}/2\phi _{0}$ being the
magnetic flux through the hexagon in units of $\phi _{0}$. $a$ is the bond
length and $k_{y}$ is the wave vector along the $y$ direction due to the
preserved translational symmetry. Note that we use the replacement $%
m^{\prime }=m-(N+1)/2$ to display the reflection symmetry of the system. And
the physical width of an armchair ribbon with width notation $N$ is $W=\sqrt{%
3}a(N-1)/2$.

For zigzag ribbons as shown in Fig. \ref{model}(d), the Harper equation reads%
\begin{equation}
E\psi _{m}=A_{m-1}^{\dag }\psi _{m-1}+C_{m}\psi _{m}+A_{m}\psi _{m+1}.
\end{equation}%
The Hamiltonian of a zigzag ribbon with width $N$ is given by
\begin{equation}
H_{zz}=%
\begin{pmatrix}
C_{1} & A_{1} & 0 & \cdots & 0 \\
A_{1}^{\dag } & C_{2} & A_{2} & \cdots & 0 \\
0 & A_{2}^{\dag } & C_{3} & \cdots & 0 \\
\cdots & \cdots & \cdots & \cdots & \cdots \\
0 & 0 & 0 & \cdots & C_{N}%
\end{pmatrix}%
,
\end{equation}%
where
\begin{align}
& A_{m}=%
\begin{pmatrix}
0 & t\cos {\varphi } & 0 \\
0 & 0 & t\sin {\varphi } \\
0 & 0 & 0%
\end{pmatrix}%
, \\
& C_{m}=%
\begin{pmatrix}
0 & \cos {\varphi }g_{3m+} & 0 \\
\cos {\varphi }g_{3m+}^{\ast } & 0 & \sin {\varphi }g_{3m-} \\
0 & \sin {\varphi }g_{3m-}^{\ast } & 0%
\end{pmatrix}%
,
\end{align}%
with $g_{3m\pm }=2t\cos [\pi f(m^{\prime }\pm \frac{1}{6})+\frac{\sqrt{3}a}{2%
}k_{y}]$. And the physical width of a zigzag ribbon with width $N$ is $%
W=3aN/2-a/2$. For zigzag ribbons, there are three types of situations for
both the left and right boundary according to the termination atom, A edged,
B edged, and C edged boundaries, respectively. The submatrices $C_{1}$, $%
A_{1}$, $C_{N}$, and $A_{N-1}$ should be modified to fit two particular
boundaries of the ribbon. The method is straightforward and the details are
not shown here for space limitation.

\begin{figure}[tbp]
\begin{center}
\includegraphics[bb=12 16 379 295, width=3.415in]{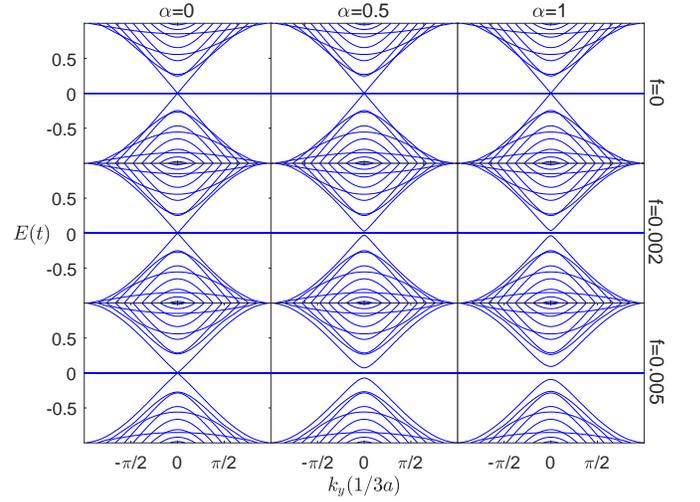}
\end{center}
\caption{Band structures of metallic armchair ribbons with various $\protect%
\alpha=0$, $0.5$, $1$ and magnetic flux $f=0$, $0.002$, $0.005$. The width
of the ribbon is fixed to $N=20$.}
\label{ekarm}
\end{figure}

\begin{figure}[tbp]
\begin{center}
\includegraphics[bb=16 1 387 295, width=3.415in]{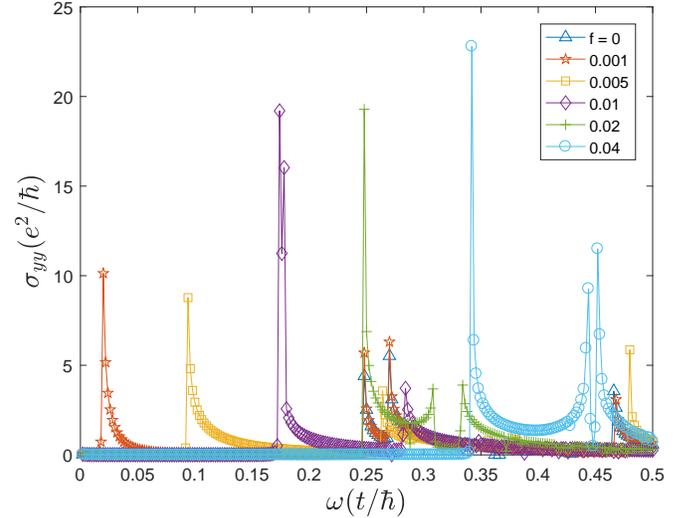}
\end{center}
\caption{Optical conductivity of a metallic armchair ribbon under various
magnetic field. The parameters are $N=20$, $\protect\alpha=1$, the chemical
potential $\protect\mu=0$, the temperature $T=300K$, and $\protect\delta%
=10^{-4}t$.}
\label{ocarm}
\end{figure}

By diagonalizing the Hamiltonian, we can obtain all the eigenvalues $%
\epsilon _{kj}$ and corresponding eigenfunctions $\psi _{kj}=\phi
_{j}e^{iky} $, where $\phi _{j}$ denotes the $j$th eigenvector of the
Hamiltonian matrix for a fixed $k_{y}=k$, with $j=1,2,3\dots 3N$. In order
to calculate the optical conductivity of ribbons using the Kubo formula, we
first write the form of the current operator along the longitudinal
direction $J_{y}=e\frac{\partial {H}}{\hbar \partial {k_{y}}}$. By
introducing the field operator $\hat{\psi}(x,y)=\sum\nolimits_{kj}\phi
_{j}e^{iky}c_{kj}$ with $c_{kj}$ being the annihilation operator in state $%
\psi _{kj}$, the current operator can be expressed in the second
quantization notation, $\hat{J}_{y}=\sum\nolimits_{kjj^{\prime
}}J_{jj^{\prime }}^{y}c_{kj}^{\dag }c_{kj^{\prime }}$, with $J_{jj^{\prime
}}^{y}=\phi _{j}^{\dag }J_{y}\phi _{j^{\prime }}$. According to the Kubo
formula, the optical conductivity is found as
\begin{equation}
\sigma _{yy}(\omega )=-\frac{1}{i\omega W}\sum_{kjj^{\prime }}\frac{%
J_{jj^{\prime }}^{y}(k)J_{j^{\prime }j}^{y}(k)(f_{kj}-f_{kj^{\prime }})}{%
\hbar \omega +\epsilon _{kj}-\epsilon _{kj^{\prime }}+i\delta },
\end{equation}%
where $f_{kj}$ is the Fermi distribution function, $\delta $ is a positive
infinitesimal, and $W$ is the physical width of the ribbon.

\section{Armchair ribbons}

\subsection{Numercial results}

Fig. \ref{ekarm} shows the evolution of the band structure of a metallic
armchair ribbon with various parameter $\alpha $ under various magnetic
field. The width of the ribbon is $N=20$, which satisfies the condition $%
N=3n-1$ of the metallic armchair ribbons. When the magnetic field is absent,
the band structure is gapless for arbitrary $\alpha$. Under a magnetic
field, the band structure opens a gap, but the flat band remains untouched.
For $\alpha=0$ (graphene), the gap is very tiny. With increasing $\alpha$,
the gap is increasing and remarkably enhanced for even modest $\alpha$.

Fig. \ref{ocarm} plots the optical conductivity of a metallic armchair
ribbon under various magnetic field. The ribbon is made from the dice
lattice ($\alpha =1$) and the width is $N=20$. Without the magnetic field,
the flat band and two linear subbands cross at zero energy, and the spectrum
is almost the same as in graphene. The first absorption peak appears at a
frequency $0.25t/\hbar $ which corresponds to the transition between the
flat band and the first parabolic subband. When the magnetic field is
present, there opens a gap in the band structure. The numerical results show
that the first response peak in the optical conductivity has the frequency
which corresponds exactly to the opened gap. This absorption peak is
attributed to the transitions between the flat band and two gapped subbands.
The gap opening remarkably reduces the threshold frequency. For $f=0.001$,
the threshold frequency is reduced to $0.02t/\hbar $. It is noted that the
gap and the threshold frequency can be continuously tuned by the magnetic
field. By comparison with the earlier results for graphene without the flat
band \cite{liu08,vacacela16}, it is clearly shown that the magnetic
modulation effect on the optical response is significantly enhanced.

For armchair ribbons with width $N\neq 3n-1$, there is a gap in the band
structure due to the discreteness of $k_{x}$. Applying a weak magnetic field
only modifies slightly the gap and shifts the first absorption peak weakly.
For comparison with graphene, $t=3$ eV and $a=1.42$ {\r{A}} are used in all
the calculations. The corresponding magnetic field is nearly $78$ T for the
magnetic flux $f=0.001$. For a realistic $\alpha -T_{3}$ material with much
larger lattice constant, this corresponding magnetic filed can be much
reduced. Moreover, it is noticeable that even a very weak magnetic field can
open a gap and induce an absorption peak, but with a very low frequency. The
room temperature $T=300K$ is used in all the calculations of optical
conductivities. Fig. \ref{ocarm} shows that the new absorption peak for $%
f=0.001$ is still sharp even at room temperature. For smaller magnetic
field, the new absorption peak with lower frequency may be broaden at high
temperature. But we argue that the temperature only affects the width of
absorption peaks, not the position of them.

\begin{figure}[tbp]
\begin{center}
\includegraphics[bb=12 16 379 295, width=3.38in]{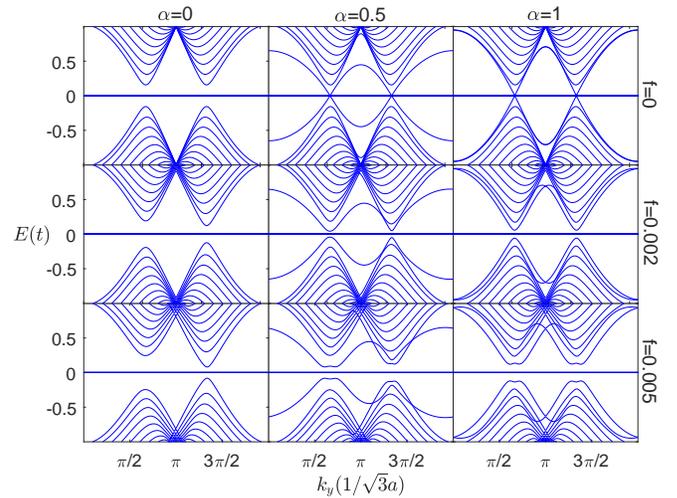}
\end{center}
\caption{Band structures of B-B edged zigzag ribbons with various $\protect%
\alpha=0$, $0.5$, $1$ and magnetic flux $f=0$, $0.002$, $0.005$. The width
of the ribbon is fixed to $N=20$.}
\label{ekzzbb}
\end{figure}

\begin{figure}[tbp]
\begin{center}
\includegraphics[bb=7 21 439 393, width=3.38in]{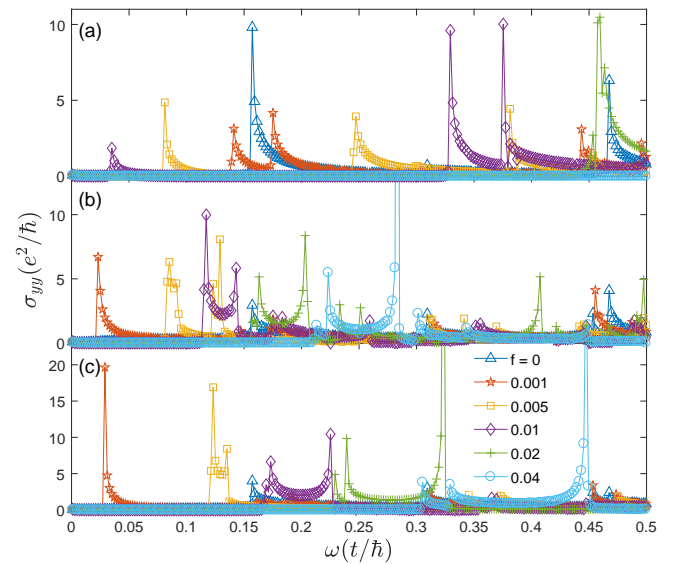}
\end{center}
\caption{Optical conductivity of a B-B edged zigzag ribbon under various
magnetic field for different parameter $\protect\alpha$: (a) $\protect\alpha%
=0$, (b) $\protect\alpha=0.5$, (c) $\protect\alpha=1$. The other parameters
are $N=20$, $\protect\mu=0$, $T=300K$, and $\protect\delta=10^{-4}t$.}
\label{oczzbb}
\end{figure}

\subsection{Analytic discussions}

To better understand the magnetic modulation effect on the optical response
in armchair ribbons made from the $\alpha -T_{3}$ lattice, we analytically
investigate the gap opening mechanism in the presence of a weak magnetic
field.

The spectrum of armchair ribbons can be obtained by imposing proper boundary
conditions to the wave functions sloved from the continuum Hamiltonian (Eq.
( \ref{hamil2}) ). For an armchair ribbon with width $N$, the boundary
conditions become $\psi (x=0)=\psi \lbrack x=\frac{\sqrt{3}a}{2}(N+1)]=0$.
Therefore the transverse momentum $k_{x}$ has to take discrete values as
\begin{equation}
k_{x}=\frac{j\pi }{\frac{\sqrt{3}a}{2}(N+1)}=\frac{2\pi }{\sqrt{3}a}\frac{l}{%
N+1},
\end{equation}%
where $l=1,2,3,\cdots N$ is the subband index. For the $l$th subband, the
wave function is
\begin{equation}
\psi _{j}=\sqrt{\frac{2}{N+1}}\xi \sin {\frac{lm\pi }{N+1}}e^{ik_{y}y}
\end{equation}%
with $m=1,2,3,\cdots N$ being the site position along the $x$ direction.
When $N=3n-1$, the $2n$th subband with $k_{x}=\frac{4\pi }{3\sqrt{3}a}\ $%
sweeps through the Dirac point $(\frac{4\pi }{3\sqrt{3}a},0)\,$, which makes
the ribbon metallic. We focus on only the flat band $\left\vert
l0\right\rangle $ (containing $N$ subbands, $l=1,2,3,\cdots N$), the $2n$th
conduction band $\left\vert 2n+\right\rangle $, and the $2n$th valence band $%
\left\vert 2n-\right\rangle $. These subbands cross at zero energy at $%
k_{y}=0$ for metallic armchair ribbons, and are crucial to the gap opening
under a magnetic field.

We consider a weak magnetic field as a perturbation. To obtain the
perturbation Hamiltonian, we first linearize $f(\boldsymbol{k})$ around the
Dirac point $\boldsymbol{K=}(\frac{4\pi }{3\sqrt{3}a},0)$ as $f(\boldsymbol{k%
})\rightarrow \hbar v_{F}(-k_{x}+ik_{y})$, where $v_{F}=\frac{3at}{2\hbar }$
is the Fermi velocity and $k_{x}$ and $k_{y}$ are measured from the $%
\boldsymbol{K}$ point from now on. And now ${\theta _{k}}=\arctan
(k_{y}/k_{x})$ becomes the momentum direction angle. Under a magnetic field,
$k_{y}$ should be replaced by $k_{y}+eA_{y}/\hbar c$ with $A_{y}=Bx$. Then
the perturbation Hamiltonian of the magnetic field reads
\begin{equation}
H_{1}=\pi tfm^{\prime }%
\begin{pmatrix}
0 & i\cos {\varphi } & 0 \\
-i\cos {\varphi } & 0 & i\sin {\varphi } \\
0 & -i\sin {\varphi } & 0%
\end{pmatrix}%
\end{equation}%
where $m^{\prime }=m-\frac{N+1}{2}$. Here we reset the midpoint of the
ribbon as the zero point in the $x$ axis. Now we calculate the magnetic
filed induced couplings among $\left\vert j0\right\rangle $, $\left\vert
2n+\right\rangle $, and $\left\vert 2n-\right\rangle $ as follows
\begin{eqnarray}
&&\left\langle 2n\pm \right\vert H_{1}\left\vert 2n\pm \right\rangle =\pm
\pi tf\sin {\theta _{k}}\sum_{m}m^{\prime }\sin ^{2}(\frac{2\pi }{3}%
m)\approx 0,  \notag \\
&&\left\langle 2n\mp \right\vert H_{1}\left\vert 2n\pm \right\rangle =\pm
i\pi tf\cos {\theta _{k}}\cos {2\varphi }\sum_{m}m^{\prime }\sin ^{2}(\frac{%
2\pi }{3}m)  \notag \\
&&\qquad \qquad \qquad \quad = 0,  \notag \\
&&\left\langle l0\right\vert H_{1}\left\vert 2n\pm \right\rangle =\pm \frac{i%
}{\sqrt{2}}\pi tf\cos {\theta _{kl}}\sin {2\varphi }  \notag \\
&&\qquad \qquad \qquad \quad \cdot \sum_{m}m^{\prime }\sin (\frac{l\pi }{N+1}%
m)\sin (\frac{2\pi }{3}m)  \notag \\
&&\qquad \qquad \approx \left\{
\begin{array}{cc}
\mp i\pi tf\cos {\theta _{kl}}\sin {2\varphi }\frac{8nl(N+1)^{2}}{\pi
^{2}(l^{2}-4n^{2})^{2}}, & l\text{ is odd}\notag \\
0, & l\text{ is even}%
\end{array}%
\right.  \notag \\
&&\qquad \left\langle l0\right\vert H_{1}\left\vert l^{\prime
}0\right\rangle =0,
\end{eqnarray}%
where ${\theta _{k}}$ (${\theta _{kl}}$) is the momentum direction angle of
the state $\left\vert 2n\pm \right\rangle $ ($\left\vert l0\right\rangle $).
From the above matrix elements, we can see that the gap opening is
attributed to the coupling between the flat band $\left\vert l0\right\rangle
$ and two linear subbands $\left\vert 2n\pm \right\rangle $. And the factor $%
\sin {2\varphi }$ shows that this coupling vanishes for graphene ($\alpha =0$%
). The gap opening in graphene is attributed to the coupling between two
linear subbands and other parabolic subbands with higher energies. The
energy difference between them makes the perturbation correction in energy
very tiny. The opened gap leads to the first response peak which is fully
tunable by the magnetic field.

\section{Zigzag ribbons}

\subsection{B-B edged zigzag ribbons}

For zigzag ribbons, there are three types of boundaries for both two edges,
namely, A edged, B edged, and C edged boundaries, respectively. We find that
only in B-B edged zigzag ribbons can the magnetic filed open a gap and give
rise to a new absorption peak. Fig. \ref{ekzzbb} shows the band structure of
the B-B edged zigzag ribbon with various $\alpha$ and magnetic field. When $%
\alpha=0$ and $f=0$, the spectrum is gaped due to the quantum confinement
effect. The flat band and the zigzag edge state subbands are degenerate at
zero energy and go through the whole Brillouin zone. With nonzero $\alpha$,
two zigzag edge state subbands become linearly dispersive subbands and cross
each other at two Dirac points. When the magnetic field is applied, the
coupling between the flat band and two linear subbands opens a gap, which is
similar to the situation in metallic armchair ribbons. It is also worthy to
note that the valley degeneracy of the band structure is lifted by the
magnetic field in the case of $\alpha<1$, which will lead to the splitting
of response peaks.

The corresponding optical conductivities are plotted in Fig. \ref{oczzbb}.
When $\alpha=0$ and $f=0$, the spectrum is gapped and the first absorption
peak is due to the transition between zero-energy edge states and the first
parabolic subband. When the magnetic field is tuned on, the peak is
splitted. With increasing magnetic field, the splitting is larger. However,
the situation is quite different in the case of $\alpha>0$. In these cases,
the zero-energy edge states evolve into two linearly dispersive subbands
around the Dirac points. But the first absorption peak changes little
because it still corresponds to the transition between the flat band to the
first parabolic subband. When the magnetic field is applied, two linear
subbands open a gap. The transitions between the flat band and two gapped
subbands give rise to a new absorption peak with much reduced frequency, as
shown in Figs. \ref{oczzbb}(b) and \ref{oczzbb}(c).

\begin{figure}[tbp]
\begin{center}
\includegraphics[bb=12 16 379 295, width=3.415in]{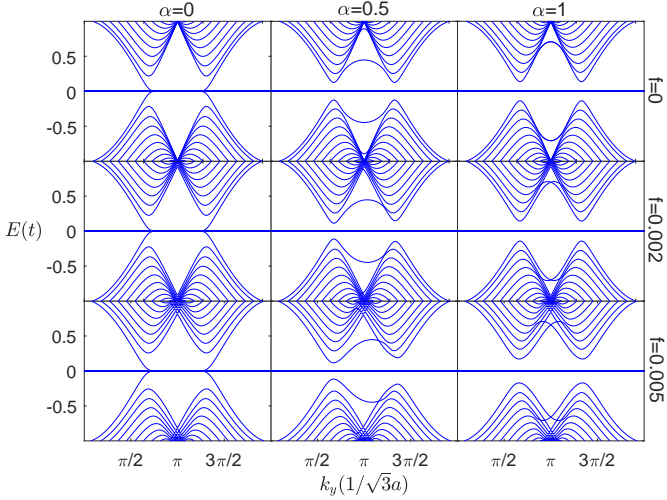}
\end{center}
\caption{Band structures of C-A edged zigzag ribbons with various $\protect%
\alpha=0$, $0.5$, $1$ and magnetic flux $f=0$, $0.002$, $0.005$. The width
of the ribbon is fixed to $N=20$.}
\label{ekzzca}
\end{figure}

\begin{figure}[tbp]
\begin{center}
\includegraphics[bb=7 21 439 393, width=3.415in]{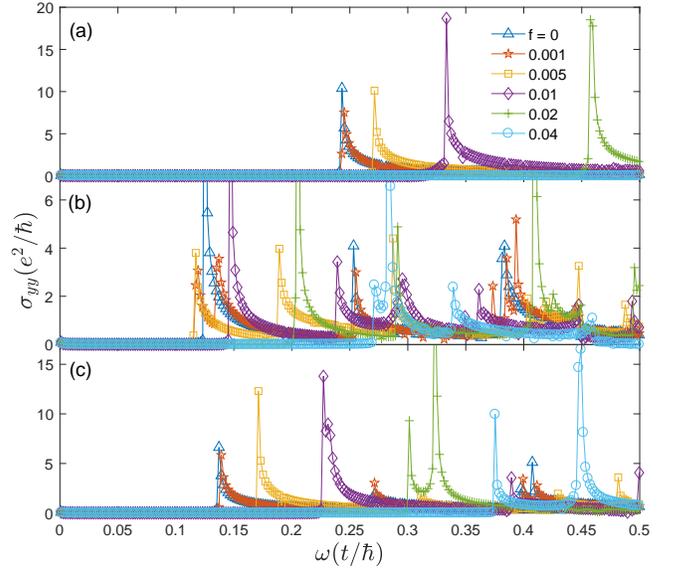}
\end{center}
\caption{Optical conductivity of a C-A edged zigzag ribbon under various
magnetic field for different parameter $\protect\alpha$: (a) $\protect\alpha%
=0$, (b) $\protect\alpha=0.5$, (c) $\protect\alpha=1$. The other parameters
are $N=20$, $\protect\mu=0$, $T=300K$, and $\protect\delta=10^{-4}t$.}
\label{oczzca}
\end{figure}

\subsection{C-A edged zigzag ribbons}

For zigzag ribbons with other boundaries, the applied magnetic field can not
give rise to a new absorption peak. Because the situation is similar for all
these zigzag ribbons, we take C-A edged zigzag ribbon as an example to
present the results.

Fig. \ref{ekzzca} shows the band structure of the C-A edged zigzag ribbon
with various $\alpha$ and magnetic field. When $\alpha=0$ and $f=0$, the
spectrum is gapless because the edge states continuously connect to the bulk
states. With nonzero $\alpha$, the edge states are gapped. An applied weak
magnetic field only slightly modifies the gap, and the modification is
different for two valleys when $0<\alpha<1$, which leads to the lifting of
valley degeneracy. In the case of $\alpha=0$, the applied magnetic field
only slightly modifies the bulk subbands and the edges states remain gapless.

The corresponding optical conductivities are plotted in Fig. \ref{oczzca}.
For $\alpha=0$, the first absorption peak is due to the transition between
zero-energy edge states and the first parabolic subband. The applied weak
magnetic field only slightly shifts the first peak. With increasing magnetic
field, the shifting is larger. For nonzero $\alpha$, the edge states are
gapped and the first absorption peak emerges due to the transition between
the flat band and the newly gapped edge states. When $\alpha=1$, the valley
degeneracy is preserved and the magnetic field only shifts the first peak.
But when $0<\alpha<1$, the valley degeneracy is lifted and then the magnetic
field splits the first absorption peak. Overall, the magnetic field can not
give rise to a new absorption peak in zigzag ribbons where the two
boundaries are not B-B edged.

\section{Conclusion}

In conclusion, we investigated the influence of a weak magnetic field on the
band structure and magneto-optical properties of nanoribbons made from the $%
\alpha-T_3$ lattice. It is found that the magnetic field can open a gap in
the band structure and induces a new absorption peak with much reduced
frequency in metallic armchair ribbons and B-B edged zigzag ribbons. The
flat band plays a key role in the gap opening and the emergence of a new
absorption peak in the optical conductivity. And this magneto-optical
modulation effect is much stronger than that in metallic armchair graphene
ribbons due to the presence of the flat band. We explained the enhancement
of the optical response under a magnetic field within the perturbation
theory for metallic armchair ribbons. And the situation is similar in B-B
edged zigzag ribbons. For the applications in magneto-optics devices, the
applied magnetic field is much weaker than that needed in graphene.
Therefore controlled terahertz radiation can be achieved much more easily in
nanoribbons made from the $\alpha-T_3$ lattice. Besides, we also find the
magnetic field will lift the valley degeneracy and split the absorption
peaks in zigzag ribbons. We propose that this enhanced magneto-optical
response may be observable in the quantum wire of critically doped Hg$_{1-x}$%
Cd$_x$Te. These findings pave the way for magneto-optics devices based on
the $\alpha-T_3$ model materials.

\begin{acknowledgments}
The work described in this paper is supported by the National Natural
Science Foundation of China (NSFC, Grant Nos. 11774144, 11774006, and
11574045), and NBRP of China (2012CB921300).
\end{acknowledgments}

\end{document}